\documentclass[aps,prl,twocolumn,showpacs]{revtex4}

\usepackage{graphicx}

\usepackage{amsmath} \newcommand{\EqLabel}[1]{\label{#1}}
\newcommand{\mb}[1]{\mathbf{#1}}

\begin{document}
\title{Holstein magneto-polarons: from Landau levels to Hofstadter butterflies}

\author{Mona Berciu}

\affiliation{ Department of Physics and Astronomy, University of
  British Columbia, Vancouver, BC, Canada, V6T~1Z1}

\begin{abstract}
We study the Holstein polaron in  transverse magnetic field using
non-perturbational methods. At strong fields and large coupling, we
show that the  polaron has a Hofstadter spectrum, however
very distorted and of lower symmetry than that of a (heavier) bare
particle. For weak magnetic fields, we identify non-perturbational
behaviour of the Landau levels not previously known.
\end{abstract}


\maketitle

The single polaron is the quintessential example of a dressed
quasiparticle: as the electron interacts with bosonic modes from its
environment, such as phonons, magnons or orbitons, it becomes
``dressed'' by a cloud of bosonic excitations. The properties of the
resulting composite object -- the polaron -- can be significantly
renormalized as compared to those of the bare particle. Accurate
numerical \cite{Holger} and analytical \cite{MA,MAo} ways to study
such problems for any strength of the electron-boson coupling and in
various dimensions have been developed in recent years.  This is to be
contrasted with the case of dressed quasiparticles in strongly
interacting systems, whose clouds consist of particle-hole
excitations. Except for the few models with known exact solutions,
their study away from perturbational regimes is still hampered by lack
of accurate and efficient methods.

Even though it is known that polarons have complex spectra, with
substantial weight up to quite high energies above the low-energy
polaron band, it is quite customary to expect that their behavior can
be understood by thinking of them as bare particles with a
renormalized mass $m^*$. In this Letter, we test this assumption by
studying the response of polarons on two-dimensional (2D) lattices to
an applied transverse magnetic field $B$.  Note that for weak
electron-phonon coupling, this problem has been studied extensively in
continuous (as opposed to lattice) models using perturbation theory,
because of its relevance to magnetotransport in 2D hetero-structures
\cite{Sarma}. For weak $B$ fields, it confirms the above-mentioned
expectation by finding that the cyclotron frequency is defined by the
polaron effective mass $m^*$. As strong fields, it predicts an
``un-dressing'' of the quasiparticle and a cyclotron frequency
controlled by the bare mass $m$. 

We use accurate non-perturbational methods to study the lattice
problem for both weak and strong electron-boson coupling.  To the best
of our knowledge, this is the first time that a polaron  lattice model
in a transverse magnetic field has ever been investigated
non-perturbationally. For weak coupling and weak fields we confirm the
results of perturbational studies at low energies. However, at higher
energies we show the emergence of a complex pattern not predicted
before, which is due to higher energy features of the polaron
spectrum.

For large magnetic fields and strong couplings, we investigate for the
first time the Hofstadter spectrum of small polarons. As is well
known, if the flux $\phi =Ba^2$ through the unit cell of a square
lattice with lattice constant $a$ is $\phi/\phi_0=p/q$, where
$\phi_0=h/e$ is the quantum of magnetic flux and $p$ and $q$ are
mutually prime integers, the free-particle band splits into $q$
sub-bands -- the  Hofstadter butterfly~\cite{Hofs}. We show that
this splitting into $q$ subbands holds for the small polaron as
well. However, the pattern is significantly distorted and of lower
symmetry than that of the bare particle, even for a coupling so large
that $m^*/m \approx 100$. This disagrees with the strong-coupling
perturbational prediction of 
 a simple mass renormalization. Taken together, these results show
 that at higher energies 
 and/or for intermediary electron-boson  couplings, the behavior of
 polarons is quite 
 different from that of bare particles with larger mass $m^*$.

{\em Model:} We investigate the Holstein model \cite{Holst} -- the
simplest and most studied lattice model of electron-phonon
interactions. The method we use is the Momentum Average (MA)
approximation, which has been shown to be highly accurate \cite{MA} not only for
this  but also for many other models, {\em eg.} with complex
lattices, $g(q)$ and $g(k,q)$ coupling \cite{MAo}, and disorder or
inhomogeneities \cite{MAd}. Here we show that MA can also treat
magnetic fields without any further approximations.  The Hamiltonian
is:
%%%%%%%%%%%%%%%%%%%%%%%%%%%%%% EQUATION %%%%%%%%%%%%%%%%%%%%%%%%%%%%%%
\begin{equation}
%\EqLabel{1}
\nonumber
{\cal H} = \sum_{\langle i, j\rangle}^{}\left[t_{ij} c^\dagger_i c_j
  + h.c.\right] +
\Omega \sum_{i}^{} b^\dagger_i b_i + g \sum_{i} c^\dagger_i c_i
\left(b^\dagger_i + b_i\right)
\end{equation}
%%%%%%%%%%%%%%%%%%%%%%%%%%%%%%%%%%%%%%%%%%%%%%%%%%%%%%%%%%%%%%%%%%%%%%
where $i$ indexes sites on a square lattice and $c_i$, $b_i$ are
electron/boson annihilation operators. The
nearest-neighbour (nn) hopping  $t_{ij} = -t \exp\left[{ie\over \hbar} \int_{j}^{i}
  \vec{A}(\vec{r}) d\vec{r}\right]$ has a Peierls phase defined by
$\vec{A}(\vec{r}) = {B\over 2}(-{y}, {x})$, 
$\Omega$ is the energy of the Einstein bosons and $g$ is the
strength of the electron-boson coupling. For $B=0$, the spin of the
electron is irrelevant and is customarily ignored. For finite $B$, the
spin degree of freedom is responsible
for a trivial Zeeman splitting between spin-up and
spin-down polaron states, which we also ignore in the following.

The quantity of interest is the Green's function:
%%%%%%%%%%%%%%%%%%%%%%%%%%%%%% EQUATION %%%%%%%%%%%%%%%%%%%%%%%%%%%%%%
\begin{equation}
\EqLabel{2}
G(i,j,\omega) = \langle 0 |c_i \hat{G}(\omega) c_j^\dagger|0\rangle =
\sum_{\alpha}^{} \frac{\langle 0 |c_i |\alpha\rangle \langle \alpha
  |c_j^\dagger|0\rangle}{\hbar \omega - E_\alpha + i \eta} 
\end{equation}
%%%%%%%%%%%%%%%%%%%%%%%%%%%%%%%%%%%%%%%%%%%%%%%%%%%%%%%%%%%%%%%%%%%%%%
where $|0\rangle$ is the vacuum, $\hat{G}(\omega) = [\hbar \omega -
  {\cal H} + i \eta]^{-1}$ is the resolvent, $\eta > 0$ is
infinitesimally small,  and the second equality
is the Lehmann representation in terms of the single-electron eigenstates ${\cal
  H} |\alpha\rangle = E_\alpha |\alpha\rangle$. In particular, we will
focus on the density of states (DOS):
%%%%%%%%%%%%%%%%%%%%%%%%%%%%%% EQUATION %%%%%%%%%%%%%%%%%%%%%%%%%%%%%%
\begin{equation}
\EqLabel{3}
\rho(\omega) = -{1\over\pi} \mbox{Im } G(i,i,\omega) =
\sum_{\alpha}^{}  |\langle 0 |c_i |\alpha\rangle |^2 \delta(\hbar
\omega - E_\alpha).
\end{equation}
%%%%%%%%%%%%%%%%%%%%%%%%%%%%%%%%%%%%%%%%%%%%%%%%%%%%%%%%%%%%%%%%%%%%%%
Because this Hamiltonian is invariant to translations,
there is no difference between local and total DOS.

The MA approach has been discussed at length elsewhere \cite{MA, MAo,
  MAd}; we review here only the salient points. We start with the
MA$^{(0)}$ formulation, which is equivalent to a variational expansion
$|\alpha\rangle = \sum_{i,j,n}^{} \phi_{i,j,n} c^\dagger_i
(b_j^\dagger)^n|0\rangle$ -- {\em i.e.} a cloud with any
number of phonons can form at arbitrary distances from the
electron, but all phonons are restricted to be at the same
site \cite{MAi}. Using this, we generate equations of motion linking
$G(i,j,\omega)$ to the generalized Green's functions $F_n(i,j,\omega)
= \langle0 | c_i \hat{G}(\omega) c_j^\dagger
(b_j^\dagger)^n|0\rangle$, as shown in Ref. \cite{MA}. The first
(exact) equation reads:
%%%%%%%%%%%%%%%%%%%%%%%%%%%%%% EQUATION %%%%%%%%%%%%%%%%%%%%%%%%%%%%%%
\begin{equation}
\EqLabel{4}
G(i,j,\omega) = G_0(i,j,\omega) + g\sum_{l}^{}F_1(i,l,\omega) G_0(l,j,\omega).
\end{equation}
%%%%%%%%%%%%%%%%%%%%%%%%%%%%%%%%%%%%%%%%%%%%%%%%%%%%%%%%%%%%%%%%%%%%%%
For any $n\ge 1$, we find within MA$^{(0)}$ that $
F_n(i,j,\omega) = gG_0(j,j,\omega-n\Omega) \left[n F_{n-1}(i,j,\omega)
  + F_{n+1}(i,j,\omega)\right]$. 
This recurrence equation is solved in terms of continuous fractions \cite{MA}
to give $F_n(i,j,\omega) = A_n(\omega) F_{n-1}(\omega)$, where
%%%%%%%%%%%%%%%%%%%%%%%%%%%%%% EQUATION %%%%%%%%%%%%%%%%%%%%%%%%%%%%%%
\begin{equation}
\EqLabel{6}
A_n(\omega) = \frac{ngG_0(j,j,\omega-n\Omega) }{1-
  gG_0(j,j,\omega-n\Omega) A_{n+1}(\omega)}
\end{equation}
%%%%%%%%%%%%%%%%%%%%%%%%%%%%%%%%%%%%%%%%%%%%%%%%%%%%%%%%%%%%%%%%%%%%%%
are independent of $j$ because ${\cal H}_0$ is invariant to
translations. Using $G(i,j,\omega) = 
F_0(i,j,\omega)$ in Eq. (\ref{4}) gives:
%%%%%%%%%%%%%%%%%%%%%%%%%%%%%% EQUATION %%%%%%%%%%%%%%%%%%%%%%%%%%%%%%
\begin{equation}
\EqLabel{7}
G(i,j,\omega) = G_0(i,j,\omega-\Sigma_{MA^{(0)}}(\omega))
\end{equation}
%%%%%%%%%%%%%%%%%%%%%%%%%%%%%%%%%%%%%%%%%%%%%%%%%%%%%%%%%%%%%%%%%%%%%%
where $\Sigma_{MA^{(0)}}(\omega) = gA_1(\omega)$. The only difference
between this and the $B=0$ solution is that here $G_0(i,j,\omega)$ is
the free-electron propagator {\em in the transverse magnetic field}.
This can be calculated efficiently as shown in Ref. \cite{new}.

While MA$^{(0)}$ is accurate in describing ground-state (GS)
properties for any effective coupling $ \lambda = {g^2/(4t \Omega)} $
so long as one avoids the extreme adiabatic limit $\Omega/t
\rightarrow 0$, it does not properly account for the polaron+one-boson
continuum that starts at $E_{GS}+\Omega$, where $E_{GS}$ is the
polaron GS energy. This feature in the spectrum is due to excited
states with a boson far away from the polaron. To properly describe
it, one needs to use MA$^{(1)}$ or a higher level \cite{MAi}. At
the MA$^{(1)}$ level, the variational basis is augmented with states
like $c^\dagger_i (b_j^\dagger)^n b^\dagger_l|0\rangle$ with $l\ne j$,
{\em i.e.}  precisely the states contributing to the continuum. The
equations of motions now also involve generalized Green's functions
related to these states, which  can be solved similarly like for the
$B=0$ case \cite{MA}. The final result is similar to Eq. (\ref{7}),
but the self-energy has the more accurate expression:
$$ \Sigma_{MA^{(1)}} (\omega)= \frac{g^2 G_0(j,j,\tilde{\omega})}{1-
  gG_0(j,j,\tilde{\omega})\left[A_2(\omega) -
    A_1(\omega-\Omega)\right] }
$$ where $ \tilde{\omega} = \omega - \Omega -
\Sigma_{MA^{(0)}}(\omega-\Omega) $. Again, the only difference from
the $B=0$ result is the $G_0(j,j,\omega)$ value. The self-energy's dependence
only on $\omega$ is due to the simplicity of the Holstein model
\cite{MA}. It becomes (weakly) non-local from the MA$^{(2)}$
level. Models with $g(q)$ and $ g(k,q)$ coupling have strong momentum
dependence in $\Sigma$ \cite{MAo}, but a finite $B$ also only requires
replacing free electron propagators with those in transverse field.

All the results shown below are for the MA$^{(1)}$ level. Like at
$B=0$, this method is also equivalent to a summation of all diagrams in the
self-energy, up to exponentially small terms discarded from each. The
resulting Green's function satisfies exactly the first 8 spectral
weight sum rules, and with good accuracy the higher order ones
\cite{MA}. While we do not know of any finite $B$ numerical results
for a direct comparison, the fact that the field is exactly included
in the free propagator together with the arguments listed above, give
us confidence that MA remains at least as accurate at finite $B$ as it
is at $B=0$ \cite{MA}.

\begin{figure}[t]
\includegraphics[width=\columnwidth]{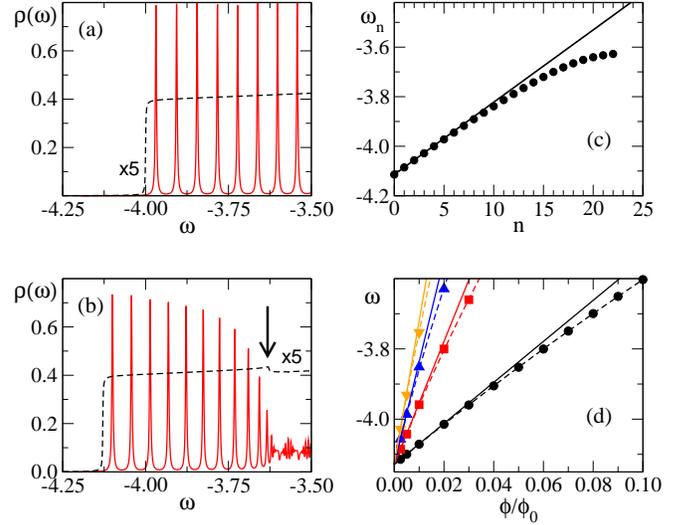}
\caption{(color online) Density of states $\rho(\omega)$
  vs. energy $\omega$, for 
  (a) $\lambda=0$ and (b) $\lambda=0.2$. Dashed lines are for $B=0$
  (times 5, for visibility), and full lines for
  $\phi/\phi_0=0.005$. The arrow in (b) marks the edge of the
  polaron+one phonon continuum. (c) Energies $\omega_n$ of the Landau
  levels vs. $n$ for $\phi/\phi_0=0.0025, \lambda=0.2$. The solid line
  is the prediction of Eq. (\ref{10}); (d) The energy of the four lowest
  Landau levels as a function of $\phi/\phi_0$ for $\lambda=0.2$. The
  lines are the predictions of Eq. (\ref{10}). Other parameters are $t=1,
  \Omega=0.5, \eta=0.002$.  }
%For fit, Z=0.928918, egs=-4.12888}
\label{fig1}
\end{figure}

{\em Results:} We begin with a weak field and weak electron-boson
coupling, where we can compare with known perturbational results
\cite{Sarma}. In Fig. \ref{fig1}(a) we plot the DOS with/without
(full/dashed line) a very small field $\phi/\phi_0= 0.005$, in the
absence of electron-boson coupling $\lambda=0$. The $B=0$ DOS is
increased 5-fold for ease of view. As expected, it has a sharp rise at
$-4t$ and then increases slowly. For $B\ne 0$, we see the Landau
levels as a succession of Lorentzian peaks with width defined by
$\eta$.

Fig. \ref{fig1}(b) shows the DOS at a weak coupling
$\lambda=0.2$. The $B=0$ band-edge has moved below $-4t$, due to the
formation of the polaron band. The top of the polaron band and the
jump marking the edge of the polaron+one-boson continuum at
$E_{GS}+\Omega$ are clearly visible (arrow). For $B\ne 0$, the
polaron band splits into LLs with smaller spacing.
Fig. \ref{fig1}(c) shows their energies $\omega_n$ when
$\phi/\phi_0=0.0025$. The line shows the
perturbational prediction: 
%%%%%%%%%%%%%%%%%%%%%%%%%%%%%% EQUATION %%%%%%%%%%%%%%%%%%%%%%%%%%%%%%
\begin{equation}
\EqLabel{10} \hbar \omega_n = E_{GS} + \hbar \omega_c^* \left( n+
        {1\over 2}\right)
\end{equation}
%%%%%%%%%%%%%%%%%%%%%%%%%%%%%%%%%%%%%%%%%%%%%%%%%%%%%%%%%%%%%%%%%%%%%%
where $\omega_c^*= eB/m^*$ is the cyclotron frequency, and we used the
$B=0$ value of polaron effective mass, $m^*$ and of $E_{GS}$. (Here, $E_{GS}=-4.1288t$
and $m^*/m=1.0765$). At low energies the agreement is very good, but
it worsens as the LLs approach the continuum. Note that as expected,
as the spacings decrease 
near the top of the polaron band, so does the spectral weight in each
LL. Even for low LLs, the agreement is worse at larger
$B$, as shown in Fig. \ref{fig1}(d). The solid lines are
Eq. (\ref{10}) using again $m^*$. The dashed lines show fits using a
$m^*(B) = m^*(1+ \gamma B)$, a correction to the polaron effective
mass predicted by perturbation theory \cite{Sarma}. The agreement is
much better although $\gamma$ increases with $n$, it is {\em not} a
constant.  Nevertheless, we
conclude that at low-energies the agreement with perturbation theory predictions is 
good \cite{note}.

At higher energies, however, it is not. Fig. \ref{fig1}(b) shows very
different DOS in the continuum than below it.  The failure of
perturbation theory here is not surprising.  Refs. \cite{Sarma} use
the free electron part as the large component, while
electron+one-boson states are the small perturbation in the
wavefunction. This is an accurate description at low energies but it
fails at the top of the polaron band and inside the continuum, where
the electron+one-boson states are dominant (for small $\lambda$) while
the free electron part is small. This failure of non-degenerate
perturbation theory at these higher energies is well known for $B=0$
models, see for example Fig. 4 of Ref. \cite{Nik}.

\begin{figure}[t]
\includegraphics[width=0.97\columnwidth]{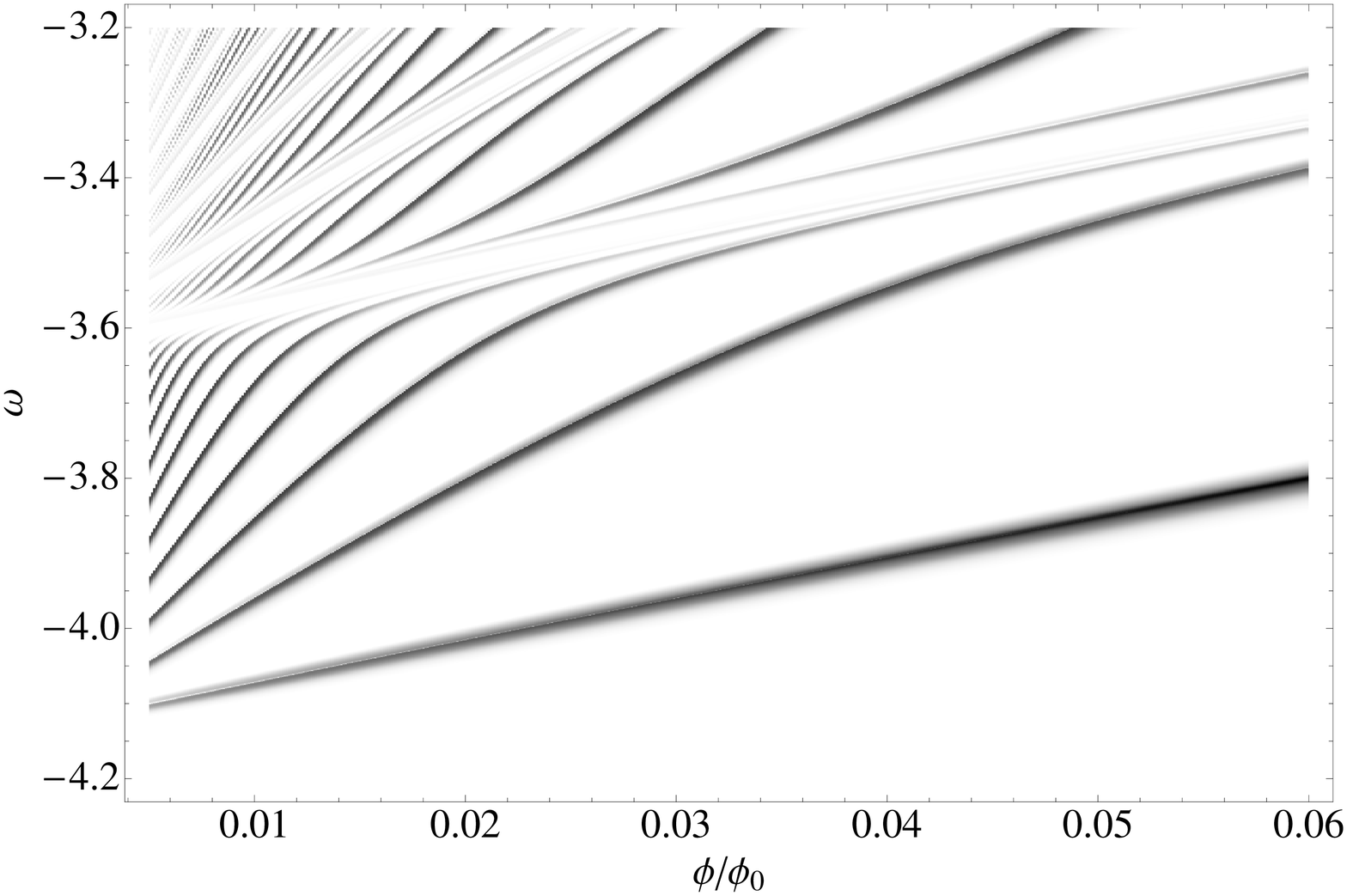}
\caption{ Contour plot of the density of states $\rho(\omega)$ vs
  $\omega$ and $\phi/\phi_0\in [0.005, 0.06]$, for $\lambda=0.2$,  $t=1,
  \Omega=0.5, \eta=0.005$.  }
%For fit, Z=0.928918, egs=-4.12888}
\label{fig2}
\end{figure}

What happens at these higher energies and also higher fields is shown
in Fig. \ref{fig2}: the polaron LLs move to higher energies as $B$
increases, until reaching an avoided crossing at an energy defined by
the continuum band-edge as $B\rightarrow 0$, and which also moves
higher with $B$. Above it, we see a whole sequence of such avoided
crossings at energies that increase faster and faster with increasing
$B$.

The reason for this beautiful spectrum is easy to find. As mentioned,
the $B=0$ polaron+one-boson continuum is due to excited states with a
boson far from the polaron. At finite $B$, the continuum splits in a
set of excited discrete states of energy $\hbar \omega_n + \Omega$, each with a
boson far from the polaron in a LL state. The DOS weights the spectrum with the 
overlap with a free particle (zero bosons) state, see Eq. (\ref{3}), so
it vanishes at these energies. This explains the sequence of avoided
crossings that occur at energies $\Omega$ above that of the low-energy LLs.

To our knowledge, this is the first time that this complex pattern is
revealed. Perturbation theory \cite{Sarma} predicts an avoided
crossing at $\hbar \omega_c \approx \Omega$, {\em i.e.} at
$\phi/\phi_0\approx 0.04$ for the values of Fig. \ref{fig2}. This is
wrong but not surprising since as mentioned, non-degenerate
perturbation theory is no longer valid at these energies.
Perturbation theory also makes predictions about high fields 
$\hbar \omega_c\gg\Omega$. Here,  Hofstadter butterfly effects
become important for our model (they are absent in Refs.~\cite{Sarma} which use
continuous models). For small $\lambda$, the DOS is quite
complex because of  overlap with the continuum and higher energy features. The
results will be discussed elsewhere. 

Instead, here we focus on another interesting question, namely
how like a particle is a strongly dressed quasiparticle?  To answer this, we 
look at the Hofstadter spectrum of a small polaron, for
$\lambda > 1$. As is well-known,  at $B=0$ the small polaron
band flattens considerably  and  a gap opens between it and the
higher-energy features~\cite{Holger, MA}. This gap allows us to look at the 
polaron response alone, avoiding
overlap with these higher energy features.

\begin{figure}[t]
\includegraphics[width=0.99\columnwidth]{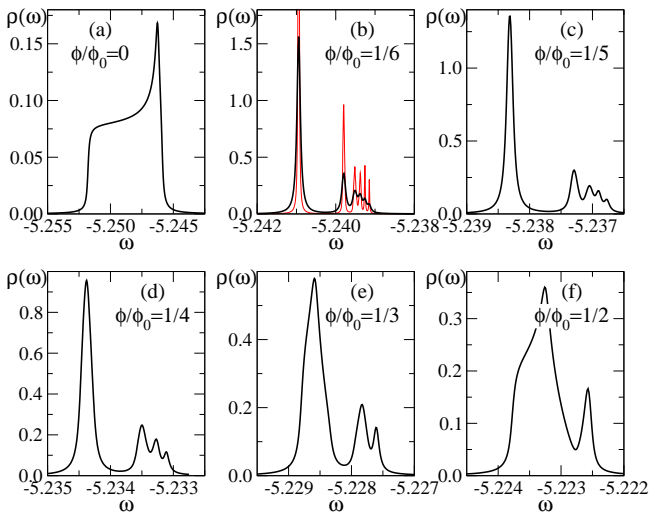}
\caption{ (color online) Density of states $\rho(\omega)$
  vs. energy $\omega$, for $\lambda=1.2$, $t=1$, $\Omega=0.5$ and various magnetic
  fluxes $\phi/\phi_0$. The thick black lines correspond to
  $\eta=5\cdot 10^{-5}$, while the thin red line in panel (b) is for $\eta
  = 10^{-5}$. The polaron band shows the Hofstadter signature,
  splitting into $q$ sub-bands if $\phi/\phi_0=p/q$. Higher energy
   features are not shown. }
\label{fig3}
\end{figure}

Fig. \ref{fig3} shows results for $\lambda=1.2$, a value just above
the crossover into the small polaron regime \cite{Holger,MA}. Panel
(a) is the polaron band DOS at $B=0$. The GS energy is significantly
lower because of the much larger binding energy, and the bandwidth is
very narrow because of the large effective mass $m^*$. As mentioned,
this band is now separated by a  gap from higher energy
features. 

Panels (b)-(f) show the low-energy DOS for a magnetic flux
$\phi/\phi_0=1/6 \rightarrow 1/2$.  The $B=0$ band indeed splits into
$q$ subbands for $\phi/\phi_0=p/q$, as expected for a bare 
particle.  For larger $q$ the subbands become narrower and a smaller
$\eta$ is needed. In panel (b) the thin line shows the DOS for
$\eta\rightarrow \eta/5$. The peaks increase much less than 5 times,
proving that these are true continua (although very narrow and thus
not yet fully converged at this $\eta$), not discrete Lorentzians. For
smaller $q$ the subbands become much wider than $\eta$ and are already
converged.

We have checked (not shown) that, as required, spectra are
unchanged if $\phi/\phi_0 \rightarrow 1 \pm \phi/\phi_0$.  We also see
that $E_{GS}(B)$ increases {\em significantly} with $\phi/\phi_0$,
reaching its maximum at 
$\phi/\phi_0=1/2$,  consistent with the Hofstadter spectrum of
a particle on a square lattice \cite{Hofs}. However, there are
also big differences. The polaron spectra are 
asymmetric: the lowest subband has most of the
weight and is quite distinct from the other subbands. This is
very unlike the Hofstadter spectrum of the bare particle on a square
lattice, which has particle-hole symmetry.

\begin{figure}[t]
\includegraphics[width=0.8\columnwidth]{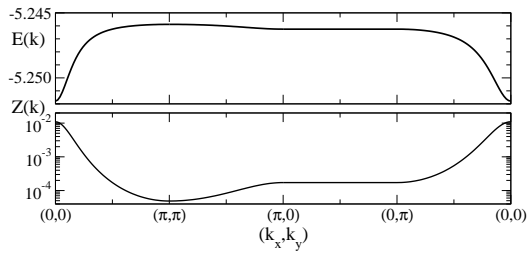}
\caption{  Polaron energy $E(\mb{k})$ (top) and
 quasiparticle weight $Z(\mb{k})$ (bottom) vs $\mb{k}$, for $B=0, t=1,
  \Omega=0.5, \lambda=1.2$. }
\label{fig4}
\end{figure}

This symmetry is lost even at $B=0$, where the DOS is not symmetric
about the center of the band. The reason (see Fig. \ref{fig4}) is that
while the van-Hove singularity is still due to the flat $E(\mb{k})$
along the $(0,\pi)-(\pi,0)$ line, it is now located just below the
upper band edge. Moreover, as shown in panel (b), the quasiparticle
weight is $\sim$ 2 orders of magnitude smaller here than near
$\mb{k}=0$. Taken together, these explain the skewed shape of the
$B=0$ DOS. They also show that nn hopping with a
$t^*/t=m/m^*=\exp(-g^2/\Omega^2)$ as predicted by first order,
strong-coupling perturbation theory~\cite{Bayo}, is not enough to fit
$E(\mb{k})$, even though $m^*/m\approx 91$. Second order perturbation
adds second and third nn hopping \cite{Bayo}:
$$ t_2^*=2t_3^* = -2{t^2\over \Omega} e^{-2{g^2\over \Omega^2}}
\sum_{n=1}^{\infty} {1\over n n!} \left({g\over \Omega}\right)^{2n}
$$ which also give a poor fit, with non-monotonic behavior along all
cuts shown in Fig. \ref{fig4}, except the $(0,\pi)-(\pi,0)$ line whose
flatness is preserved.  Indeed, to reasonably fit $E(\mb{k})$ one
needs to add $\cos(nk_x a+ mk_ya)$ terms with up to $|n|+|m|\approx
6$. In other words, one needs to include terms at least up to 6th
order in perturbation theory in the hopping Hamiltonian to properly
describe it.

The long range hopping in $E(\mb{k})$ and the varying $Z(\mb{k})$
explain the asymmetry of the polaron  Hofstadter spectra.  Taken
together with the low-$\lambda$ results, they also show that a polaron
is not behaving just like a bare heavier particle with 
mass $m^*$. Instead, its composite structure and the existence of higher
energy states signal their existence in its finite-$B$ response. This
has obvious implications for the interpretation of experimental data.

{\bf Acknowledgements:} Work supported by NSERC and CIFAR. We thank
P. Stamp for suggesting the problem, and H. Fehske and G.  Sawatzky
for discussions.


\begin{thebibliography}{99}
\bibitem{note} The work in Refs. \cite{Sarma} is for the Fr\"ohlich
 parabolic  model, however except for changed coefficients one
 expects similar results for the Holstein model at low energies.
\end{thebibliography}
\end{document}